\documentclass [10pt]{article}
\usepackage[utf8]{inputenc}
\usepackage{amsmath}
\usepackage{amsfonts}
\usepackage{amssymb}
\usepackage{graphicx}
\usepackage{soul}
\usepackage{parskip}
\usepackage{natbib}
\usepackage{bbm}
\usepackage{algorithm}
\usepackage{bm}
\usepackage[noend]{algpseudocode}
\usepackage{booktabs}
\usepackage[normalem]{ulem}

\usepackage[dvipsnames]{xcolor}

\title{\textbf{Urban Epidemic Hazard Index for Chinese Cities: Why Did Small Cities Become Epidemic Hotspots?}} 

\author{%
\textsc{Tianyi Li}\thanks{tianyil@mit.edu} \\[1ex] 
System Dynamics Group, Sloan School of Management, MIT \\ 
\\
\textsc{Jiawen Luo}\thanks{jiawen.luo@erdw.ethz.ch} \thanks{T. L and J. L contribute equally to this work.}\\[1ex] 
Institute of Geophysics, ETH Zurich \\ 
\\
\textsc{Cunrui Huang}\thanks{huangcr@mail.sysu.edu.cn} \\[1ex] 
School of Public Health, Sun Yat-sen University \\ 
\\
}
\date{\today} 

\begin{document}

\maketitle

\begin{abstract}
Multiple small- to middle-scale cities, mostly located in northern China, became epidemic hotspots during the second wave of the spread of COVID-19 in early 2021. Despite qualitative discussions of potential social-economic causes, it remains unclear how this pattern could be accounted for from a quantitative approach. Through the development of an urban epidemic hazard index (EpiRank), we came up with a mathematical explanation for this phenomenon. The index is constructed from epidemic simulations on a multi-layer transportation network model on top of local SEIR transmission dynamics, which characterizes intra- and inter-city compartment population flow with a detailed mathematical description. Essentially, we argue that these highlighted cities possess greater epidemic hazards due to the combined effect of large regional population and small inter-city transportation. The proposed index, dynamic and applicable to different epidemic settings, could be a useful indicator for the risk assessment and response planning of urban epidemic hazards in China; the model framework is modularized and can be adapted for other nations without much difficulty.
\end{abstract}

Despite the nation-wide successful implementation of control measures against COVID-19 \citep{Kret2020, Laet2020, Tiet2020}, multiple small- to middle-scale cities in China became local epidemic hotspots during the new wave of the pandemic in early 2021; the list includes Tonghua, Songyuan, Suihua, Qiqihar, Heihe, and Xingtai etc. \citep{Y2021,St2021}. Unlike their nearby capital cities (e.g., Shijiazhuang, Changchun, Harbin), these small cities are probably not known to many Chinese before they are enlisted as ``high-risk regions'' after the local bursts, and it is indeed an unexpected phenomenon that they are highlighted among the over 340 prefectural administrations in China. It is implied that the likelihood of epidemic hazards is high in these regions. Many social-economic factors may account for this fact \citep{Niuet2020,Qiuet2020}: for example, social scientists may observe that these cities are all located in the northeast part of China, where local economies are often underdeveloped, and local residents are often more behavioral active than they are supposed to be in face of the epidemic \citep[e.g.,][]{S2006}; other conjectures may attend to the fact that since these are neither coastal cities nor metropolitans where imported cases are more common, local control measures and regulations are different and somewhat relaxed, which led to heedlessness of early signals \citep[e.g.,][]{Yeet2020,Liet2021}, or that these northern regions have cold winters and also less residential housing space than the south, hence the hazard of severe infections was harbored \citep[e.g.,][]{Caoet2020}. Although these arguments are sound, it is desired that quantitative explanations could be addressed to decipher why such small cities in China stand out as epidemic hotspots. 

This points to the significant necessity of the quantification of urban regions' epidemic hazards, desirably via a constructed index that assesses the extent of potential risk. Such indices of potential epidemic risk can be useful for effective decision analysis during epidemic response and planning \citep{Met2012, Het2018}, critical to the mitigation of sudden and potentially catastrophic impacts of infectious diseases on society \citep{Set2020}. Indeed, although the exit of COVID-19 is still on the fly, both methodologically and practically \citep{Thet2020}, it is nevertheless prudent to start getting prepared for the next pandemic \citep{Let2018, NK2019, Siet2020}, both economically and ecologically \citep{Diet2020,Doet2020}, through comprehensive pandemic risk management synthesis \citep{SP2020} and the upgraded implementation of digital technologies \citep{Buddet2020} 

Many successful attempts have been made to develop such epidemic risk indices from various angles. According to US CDC, the preparedness for influenza pandemics can be assessed with the Influenza Risk Assessment Tool \citep{Tet2012, Cet2014, BT2018} through the Pandemic Severity Assessment Framework \citep{Ret2013,Hoet2014}; for the same purpose, the Tool for Influenza Pandemic Risk Assessment \citep{TIPRA} is recommended by WHO; and there are miscellaneous other tools developed for national-level pandemic planning, through either mathematical simulations \citep[e.g.,][]{Eet2007} or scoring systems, in which case various social-economic factors are being considered \citep[e.g.,][]{Met2019, Oet2019, Get2020}.

However, it is suggested that currently a suitable public health evaluation framework for the assessment of epidemic risk and response is still not full-fledge in scale \citep{Wet2020}. Most proposed tools and frameworks are subject to a few shortcomings: (1) assessments are in most cases from the supply side (i.e., the preparedness) instead of the demand side (i.e., the actual risk); (2) assessment of pandemic potential is often virus-specific (i.e., pathological), not sufficiently general-purpose, as the risk potential is also determined by various societal factors (transportation, population \citep{CG2020} etc.); (3) many indices rely on expert scoring systems that often depend largely on subjectivity, and the calling for mathematical models and algorithms for risk assessment and pandemic planning is compelling \citep{WC2011}; (4) finally, many models focus on nation-wide evaluation, and there is relatively little concentration on sub-nation (e.g., city) level analysis, except for a few interesting attempts \citep[e.g.,][]{Pret2020,BK2020,Zhuet2020}. 

To deal with this problem, in this study we develop a novel epidemic hazard index for Chinese cities, which quantifies the potential risk of epidemic spread at the over 340 Chinese prefectural administrations. The index relies on a simulation model which integrates intra-city compartment dynamics and a detailed mathematical description of inter-city multi-channel transportation. Calculation of the hazard index is based on this dynamic system that simulates the domestic epidemic spread for user-specified diseases. In the model, intra-city evolution is governed by the SEIR dynamics assuming no epidemic response taken place, such that the constructed index serves as an early-warning indicator at early periods of epidemics, before the incidence of any structural change in the population flow upon policy intervention \citep[e.g.,][]{Schlosseret2020,LiR2021}; inter-city transportation is modeled with a multi-layer bipartite network \citep{L2020}, which make explicit considerations of various events during inter-city population flow, including transit, cross-infection due to path overlap, as well as the different transmissivities on different transportation media. 

Such a highlight on transportation (i.e., spacial patterns) is core to the city-specific risk assessment of epidemics \citep{Set2002} and natural hazards in general \citep{Get2006}. Indeed, over the course of the still on-going pandemic, it is acknowledged that transportation, at both the global and the domestic level, plays a critical role during the spread of viruses and to a large extent may determine the severity of the disease at different geological divisions \citep{CC2020,Let2021}. Essentially, compared to regression models \citep{Boet2014} or the machine-learning approach \citep{Fet2021}, the highlight of transportation asks for epidemic risk analysis from a network perspective \citep[e.g.,][]{Cet2020,Geet2020,Set2020a}, upon which the risk scores could then be computed from quantitative approaches \citep{Set2020b}. An important precedent is the Global Epidemic and Mobility (GLEaM) model, which integrates sociodemographic and population mobility data in a spatially structured stochastic disease approach to simulate the spread of epidemics at the worldwide scale \citep{Bet2010}. GLEaM considers the commuting on the airport network on top of local disease transmission, where transportation is modeled via an effective operator; our model adopts a similar methodology yet constructs a more realistic multi-layer mathematical description of inter-city transportation, comparing also to various recent studies in the same line of research \citep{Jiaet2020,Wanget2020,Zhanget2020,Changet2021}.

\section*{Model}
The model is developed in \citet{L2020} and is briefed summarized here. Assume a bi-partite graph with cities (nodes) classified as either central cities or peripheral cities. The network is multi-layer $G = (V,E_{A/B/R/S})$, specifying four means of inter-city transportation (i.e., different layers have different connectivites between nodes): Air (A), Bus (B), Rail (R) and Sail (S). At each node, the local urban population is divided into four compartments: Susceptible (S), Exposed (E), Infected (I), Recovered (R), and the intra-city epidemic spread follows the standard SEIR dynamics \citep[e.g.,][]{S2000,N2010}. We track the in- and out-flow of the exposed, susceptible and recovered population at each node on the transportation network, which determines the open-system SEIR dynamics (for a specific city $i$):
\begin{equation}
\left\{
\begin{aligned}
\dot{S}_i & = - \frac{S_i}{P_i}(\frac{R_0}{D_I}I_i + z_i)+ \Delta S_i^{in}  - \Delta S_i^{out}\\
\dot{E}_i  & = \frac{S_i}{P_i}(\frac{R_0}{D_I}I_i + z_i) - \frac{E_i}{D_E}  + \Delta E_i^{in}  - \Delta E_i^{out}\\
\dot{I}_i  & = \frac{E_i}{D_E} - \frac{I_i}{D_I} \\
\dot{R}_i & = \frac{I_i}{D_I} + \Delta R_i^{in} - \Delta R_i^{out}.
\end{aligned}
\right.
\end{equation} 
Epidemiological parameters $R_0$, $D_E$, $D_I$ are the basic reproduction number, the incubation period, the infection period; $z_i=z_i(t)$ is the zoonotic force, i.e., the seed of the disease, with $z_i\neq 0$ only at node(s) that is the disease epicenter(s). The in- and out-flows of each city are determined via the flowmaps $F^q = \{f_{i,j}^q\}$ between pairs of cities $(i,j)$ specified for each means of transportation $q$. During the inter-city flow, transit events are considered, where only a certain proportion of flow ($TR_{c/p}$ for central/peripheral cities, the same for each layer) enter the local population, and the rest are directed to other destinations. Moreover, cross-infections during inter-city travels are modeled, which take place between a susceptible person and an exposed person who share an overlapped travel path with the same destination. The strength of the cross-infection spillover effect $R_T^q$ various on different transportation media \citep[e.g.,][]{Het2020}. 

At each city $i$, the exposed inflow ($\Delta E_i^{in}(t)$) is given by
\begin{equation}
\Delta E_i^{in}(t) = \underset{q\in{Q}}{\sum} \underset{j\in{V}}{\sum} \overline{f_{j,i}^q (t) \mu_j (t)} (1-TR_i),
\end{equation}  
where
\begin{equation}
\overline{f_{j,i}^q  \mu_j}= f_{j,i}^q  \mu_j + \underset{k}{\overset{p^q(i,k)\cap p^q(i,j)\neq 0}{\sum}} f_{k,i}^q \mu_k(R_T^q -1)\frac{f_{j,i}^q(1-\mu_j-\eta_j)\text{min}(d^q_{j,i},d^q_{k,i})}{\underset{l}{\overset{p^q(i,k)\cap p^q(i,l)\neq 0}{\sum}} f_{l,i}^q (1-\mu_l-\eta_l)\text{min}(d^q_{l,i},d^q_{k,i})}
\end{equation}
is the adjusted exposed flow from city $j$ to $i$ by means $q$, taking care of cross-infections ($d^q_{i,j}$ represents the shortest path distance between $i$ and $j$ on layer $q$), and
\begin{equation}
\label{eq:mu}
\mu_i(t) = \frac{\Delta E_i^{out}(t) + \underset{q\in{Q}}{\sum} \underset{j\in{V}}{\sum} \overline{f_{j,i}^q (t-1) \mu_j (t)} TR_i^q}{\underset{q\in{Q}}{\sum} \underset{j\in{V}}{\sum} f_{i,j}^q (t)},
\end{equation}
\begin{equation}
\label{eq:eta}
\eta_i(t) = \frac{\Delta R_i^{out}(t) + \underset{q\in{Q}}{\sum} \underset{j\in{V}}{\sum} f_{j,i}^q (t) \eta_j(t-1) TR_i^q}{\underset{q\in{Q}}{\sum} \underset{j\in{V}}{\sum} f_{i,j}^q (t)},
\end{equation}
are the time-stamped proportion of the exposed and recovered population among the total outflow population from city $i$; thus $(1-\mu_i-\eta_i)$ is the proportion of the susceptible population among the total outflow from city $i$. The recovered inflow is tracking all the recovered people upon arrival (via $\eta_j$): $\Delta R_i^{in}(t) = \underset{q\in{Q}}{\sum} \underset{j\in{V}}{\sum} f_{j,i}^q (t) (1-TR_i^q) \eta_j(t-1)$, and according to flow balance, the susceptible inflow is $\Delta S_i^{in}(t) = \underset{q\in{Q}}{\sum} \underset{j\in{V}}{\sum} f_{j,i}^q (t) (1-TR_i^q) - \Delta E_i^{in}(t) - \Delta R_i^{in}(t)$. The outflow population from city $i$'s population $P_i$ is the total outbound flow minus the transferred inbound flow, contributed by the $S,E,R$ compartments (with the proper assumption that $I$ stay local, i.e., infected people do not participate in inter-city travels). Proportionally, the outflows are:
\begin{equation}
\Delta X_i^{out}(t) = X_i(t) \frac{ \underset{q\in{Q}}{\sum} \underset{j\in{V}}{\sum} f_{i,j}^q (t) - \underset{q\in{Q}}{\sum} \underset{j\in{V}}{\sum} f_{j,i}^q (t) TR_i^q}{S_i(t) + E_i(t) + R_i(t)}.
\end{equation}  
with $X$ being $S$, $E$ or $R$.

\begin{figure}[h]
\centering   
	\includegraphics[width=6in]{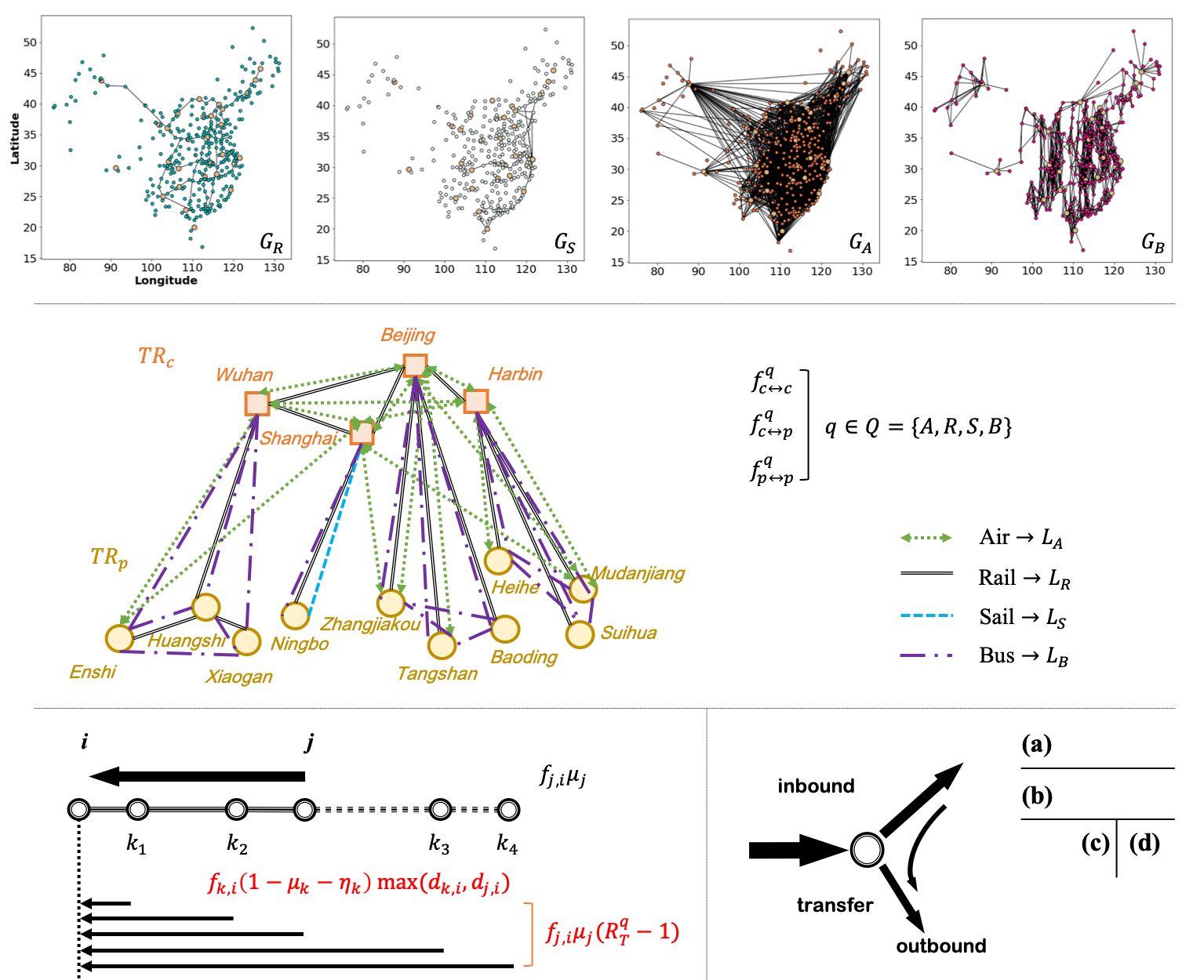}  
\caption{Multi-channel transportation network model with detailed flow description. (a) Four layers of inter-city transportation (Air, Bus, Sail, Rail) between Chinese prefectural administrations. (b) Bi-partite categorization of nodes (central cities vs. peripheral cities). (c) Cross-infection during travel due to path overlaps. (d) Inbound/outbound flow upon city arrival.} 
\label{Fig1}
\end{figure}
Overall, the multi-layer network model is summarized in Figure 1 (see more details in \citep{L2020}).

\section*{EpiRank}

With the constructed model framework, essentially a simulator of the spread of epidemics, we are able to simulate imaginary diseases of arbitrary epidemiological features originated from arbitrary epicenters. Suppose an epidemic initiated at node $i$, with a certain set of epidemiological parameters $R_0,D_E,D_I$. It is important to quantify the intensity of this epidemic, i.e.,  the extent of its spread on the domestic scale. This points to a centrality measure of the node that characterizes nodes' ability in spreading an epidemic. Borrowing from the idea of PageRank \citep{Pet1999, LM2011}, we construct this new centrality score in a way similar to the eigenvalue centrality and term it as the \textit{EpiRank}. 

Start the simulator with the disease seeded at node $i$ upon a given zoonotic force, specified with $z(t)$ during period $t_z^s$ to $t_z^e$. Consider a constant force $z_0$ over such a period (whose length is $\Delta t_z$), and the overall zoonotic force $Z$ is:
\begin{equation}
Z = \int_{t_z^s}^{t_z^e} z(t) dt \sim \sum_{t_z^s}^{t_z^e} z(t) = (t_z^e - t_z^s)z_0 = \Delta t_z z_0.
\end{equation}
The simulation proceeds, during which the local disease spreads to the entire nation via the transportation network. After $\tau$ time steps, we obtain the number of infected cases at city $j$, denoted as $I_{ij}^{\tau}$, with the first subscript indicating the epicenter. Similarly we  obtain $R_{ij}^{\tau}$, $E_{ij}^{\tau}$ etc. We define the normalized total infection at city $j$ as $U_{ij}^{\tau} = (I_{ij}^{\tau}+R_{ij}^{\tau})/Z$. $U_{ij}^{\tau}$ is then used to compute the epidemic score EpiRank for node $i$:
\begin{equation}
h_i^{\tau} = \alpha \underset{j\in{G},j\neq i}{\sum}\frac{f(U_{ij}^{\tau})}{\underset{j\in{G}}{\sum}f(U_{ij}^{\tau})} h_j^{\tau} + (1-\alpha)U_{ii}^{\tau}.
\end{equation}
$f(U_{ij}^{\tau})$ represents a specific function of $U_{ij}^{\tau}$ to indicate the relative weights of each city $j$ contributing to the score for city $i$. Here we simply allow $f(U_{ij}^{\tau}) = U_{ij}^{\tau}$ but further considerations could be made, for example, applying a cutoff $f(U_{ij}^{\tau}) = max(U_{ij}^{\tau}-U_0,0)$. 

This score $h_i$ thus indicates the spreading intensity of epidemics at any node $i$, which is contributed by (1) the city's local severity of the epidemic, and (2) its ability of spreading the disease to other cities, with the intensity scores of other cities contributing to its own score at particular weights. $\alpha\in [0,1]$ is the modulating parameter weighing over these two effects: a small $\alpha$ concerns more on the city-wide local spread of the epidemic ($\alpha = 0$ corresponds to a complete local index), while a large $\alpha$ puts more weight on the city's capability of spreading the disease out. 

Write $\bm{W} = \{W_{ij}\}= \{U_{ij}^{\tau}/\underset{j\in{G}}{\sum}U_{ij}^{\tau}\}$, then $\bm{h} = \{h_i^{\tau}\}$ could be calculated with:
\begin{equation}
\left\{
\begin{aligned}
& h_i^{\tau} = \alpha \underset{j\in{G},j\neq i}{\sum} \frac{U_{ij}^{\tau}}{\underset{j\in{G}}{\sum}U_{ij}^{\tau}} h_j^{\tau} + (1-\alpha)U_{ii}^{\tau} \\
&\Longrightarrow (1+ \alpha \frac{U_{ii}^{\tau}}{\underset{j\in{G}}{\sum}U_{ij}^{\tau}} )h_i^{\tau} = \alpha \underset{j\in{G}}{\sum} \frac{U_{ij}^{\tau}}{\underset{j\in{G}}{\sum}U_{ij}^{\tau}} h_j^{\tau} + (1-\alpha)U_{ii}^{\tau} \\
&\Longrightarrow (1+ \alpha W_{ii})h_i^{\tau} = \alpha \underset{j\in{G}}{\sum} W_{ij} h_j^{\tau} + (1-\alpha)U_{ii}^{\tau} \\
&\Longrightarrow [\bm{I} -\alpha(\bm{W} - \text{diag}(\bm{W}))]\bm{h} = (1-\alpha)\text{diag}(\bm{U}) \mathbbm{1} \\
&\Longrightarrow \bm{h} = (1-\alpha)[\bm{I} -\alpha(\bm{W} - \text{diag}(\bm{W}))]^{-1}\text{diag}(\bm{U}) \mathbbm{1}. \\
\end{aligned}
\right.
\end{equation} 

When $\alpha = 0$, $\bm{h} = \text{diag}(\bm{U}) \mathbbm{1}$. When $\alpha = 1$, $\bm{h}$ is the eigenvector of [$\bm{W} - \text{diag}(\bm{W})$] of eigenvalue 1, in which case we may impose a value for $\text{max}(\bm{h})$; wlog, we consider $\alpha \neq 1$. Note $\bm{h} = \bm{h}(\Delta t_z, z_0)$ but not $\bm{h} = \bm{h}(Z)$, since $\bm{U} = \bm{U}(\Delta t_z, z_0) \neq \bm{U} (Z)$, i.e., the spread patterns are different under different distributions of the same overall zoonotic force.

\section*{Results}

Connectivities of each layer (i.e., transportation routes) are determined from public datasets and empirical considerations; city information (population, GDP etc.) is obtained from public datasets; transportation parameters (flowmap, transfer rate, cross-infection strength) are determined through fitting the early spread of COVID-19 in China in January-February 2020, where the multi-parameter inversion is conducted via a smart gradient method (see \citep{L2020} and Supplemental Materials). Transfer rate at central/peripheral cities are $TR_c = 0.4$ and $TR_p = 0.05$; the cross-infection strength are $\{R_T^A,R_T^R,R_T^S,R_T^B\} = \{1.2,1.5,1.5, 3\}$ on the four transportation media; inter-city flows are different for different types of city pairs (central-central, central-peripheral, peripheral-peripheral), and are determined at $f_{cc/cp/pp}^A = 1000/500/0$, $\hat{f}_{cc/cp/pp}^R = 2000/200/500$, $f_{cc/cp/pp}^S = 100/100/100$, $f_{cc/cp/pp}^B = 0/3000/1000$. These transportation parameters well fit the early spread of COVID-19 in China; they are independent of epidemiological concerns and are fixed throughout the simulations. 

For epidemiological parameters, the zoonotic force $z$ is assumed to be 5 persons/day at Day 1 and zero afterwards, at a single epicenter (the simulator nevertheless allows for simultaneous bursts at multiple epicenters). The base-case disease is fixed at $R_0 = 2.5$, $D_E = 6\ days$ and $D_I = 3\ days$, i.e., a mild reproduction of virus and a medium-range infection duration, close to the clinical parameters of COVID or SARS \citep[e.g.,][]{Wuet2020}. Unintervened spread of this seeded disease to all Chinese prefectural districts is simulated for 30 days, after which the ever-infected population ($I_i+R_i$) in each city $i$ are recorded and are used to compute the hazard index $h$. 

\begin{figure}[h]
\centering   
	\includegraphics[width=6in]{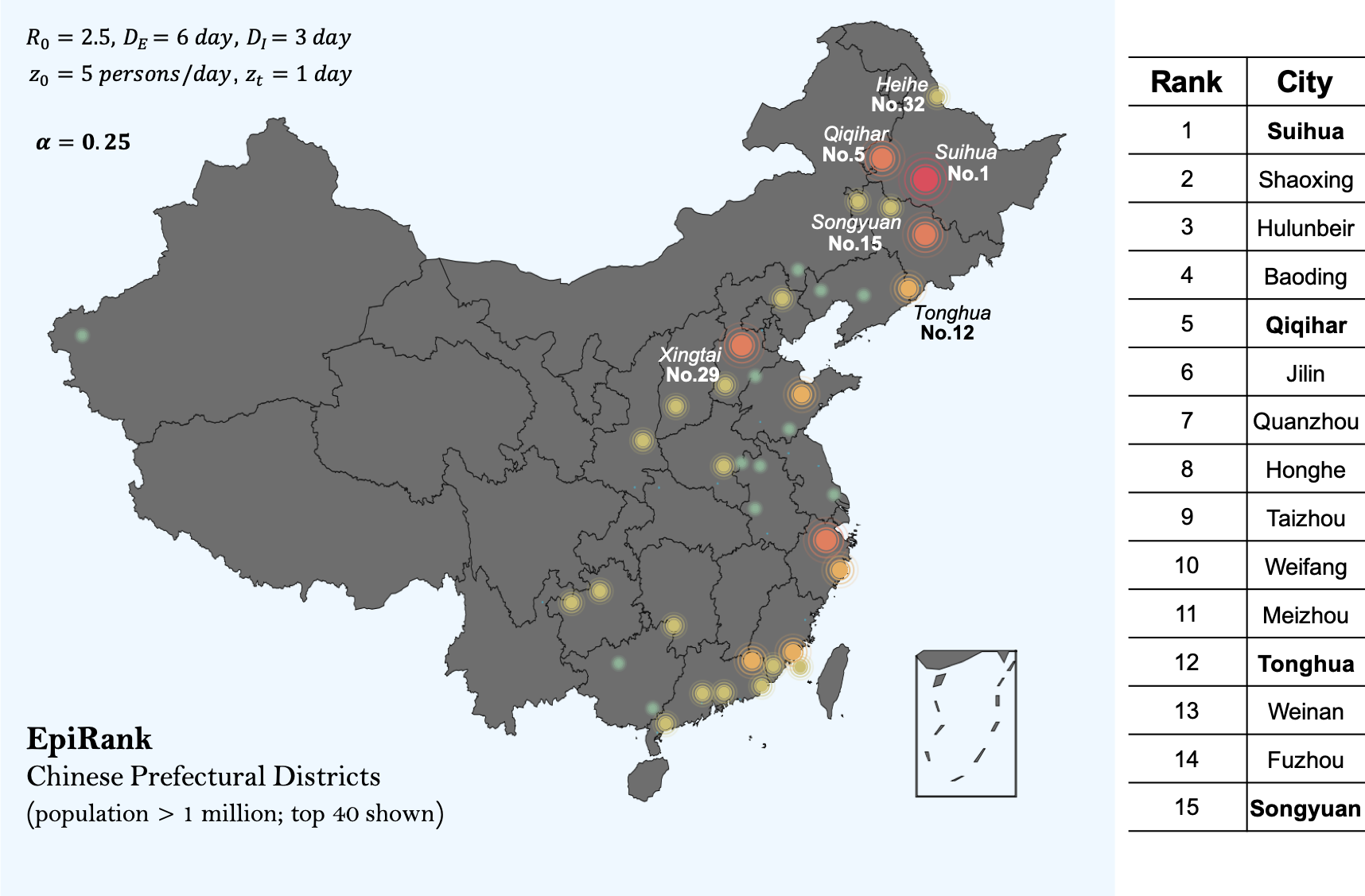}  
\caption{Urban Epidemic Hazard Index (EpiRank) of Chinese Cities (with population greater than 1 million). Assuming a simulated disease with $R_0 = 2.5$, $D_E = 6\ days$, $D_I = 3\ days$, seeded by a 1-day zoonotic force at the strength of $5\ persons/day$; $\alpha=0.25$. Table: Top 15 ranks. Among 300 cities, the six ground-true cities are successfully indicated at high rankings (four in top 5$\%$ and all six in top 10$\%$).} 
\label{Fig2}
\end{figure}

Cities having population larger than 1 million (300 out of 347) participate in the calculation of $\bm{h}$ under equation (9). For a value of $\alpha=0.25$, the determined rankings of cities' epidemic hazards are shown in Figure 2 (top 40 in the graph and top 15 in the table). One sees that, quite strikingly, the six small cities where the new bursts of COVID took place (Tonghua, Songyuan, Suihua, Qiqihar, Heihe, Xingtai) are successfully highlighted by the computed hazard index. All six cities rank within or near top 10$\%$ in the list, including four cities ranking within the top 15. Tests suggest that the result is robust; the high ranks of the six denoted ground-true cities are largely invariant to fluctuations in both transportation parameters and epidemiological parameters.

The hazard rankings are computed at different $\alpha$, and correlation of the ranks is demonstrated via the Spearman's correlation coefficient (Figure 3; comparing top 30 entries of each rank). A stable ranking at small values of $\alpha$ is identified, along with a second invariance at the larger end (Figure 3 center). Indeed, the ranking is almost completely different at, for example, $\alpha = 0.1$ vs. $\alpha = 0.8$, with the latter having a new set of cities ranked top in the list which are mostly located in the middle of China. This is consistent with our theory, as a small/large $\alpha$ points to either of the two end-members of epidemic hazards: the city-wide local spread, or the capability of spreading the disease to other locations. Therefore, cities ranking high at small $\alpha$ are regions with mass population but relatively small transportation means, in which case local epidemic bursts are severely harbored \citep{Iet2020} but not much spillovered to other cities. On the opposite, cities ranking high at large $\alpha$ are regions where inter-city transportation is sufficiently viable with respect to the humble local population; in this case, when seeded a virus, the city is less likely to become a closed epidemic cluster than to enormously spread the disease out to other regions. The epidemic hazard index thus also implies the condition of a city's inter-city transportation infrastructures.

\begin{figure}[h]
\centering   
	\includegraphics[width=5in]{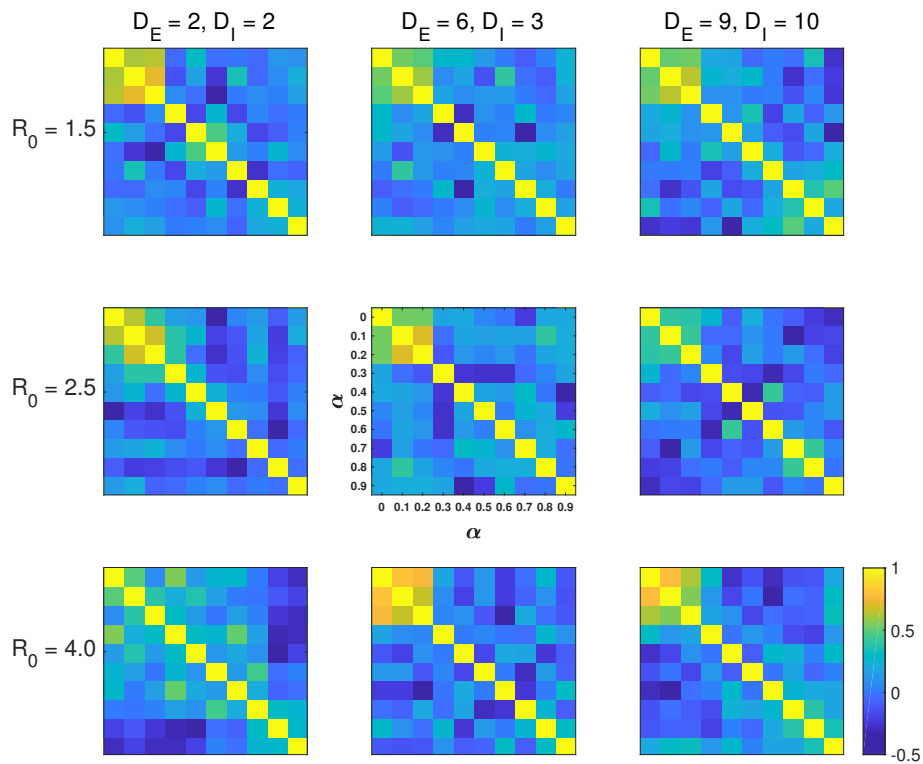}  
\caption{Spearman's correlation of hazard rankings (top 30 in the list) at different $\alpha$ (each sub-figure), for different sets of epidemiological parameters: reproduction number $R_0 = 1.5/2.5/4$ and infection duration $(D_E, D_I) = (2,2)/(6,3)/(9,10)$ days. Base case (center sub-figure) is $R_0 = 2.5$ and $(D_E, D_I) = (6,3)$.} 
\label{Fig3}
\end{figure}

We initiated simulations for different sets of epidemiological parameters of the assumed disease (Figure 3), with the combination of low/medium/high infectivity ($R_0 = 1.5/2.5/4$) and short/medium/long infection duration ($(D_E, D_I) = (2,2)/(6,3)/(9,10)$ days). The invariance at small $\alpha$ is largely maintained across the experiments, expect for a very severe virus with high infectivity and short duration ($R_0 = 4.0, (D_E, D_I) = (2,2)$). In this case, 30 days is sufficient for most population in most cities to get infected, and thus top rankings lean instead on densely populated cities. The second invariance around high $\alpha$ is also identifiable, although not as clear as the first one. In some cases there is a third cluster at intermediate values of $\alpha$, but its significance is not as high as the first two which have well-grounded interpretations. Overall, one is able to conclude that the two end-options of the epidemic hazard index, using small or large $\alpha$, hold meaningful across different scenarios of epidemic onset.

\begin{figure}[h]
\centering   
	\includegraphics[width=7in]{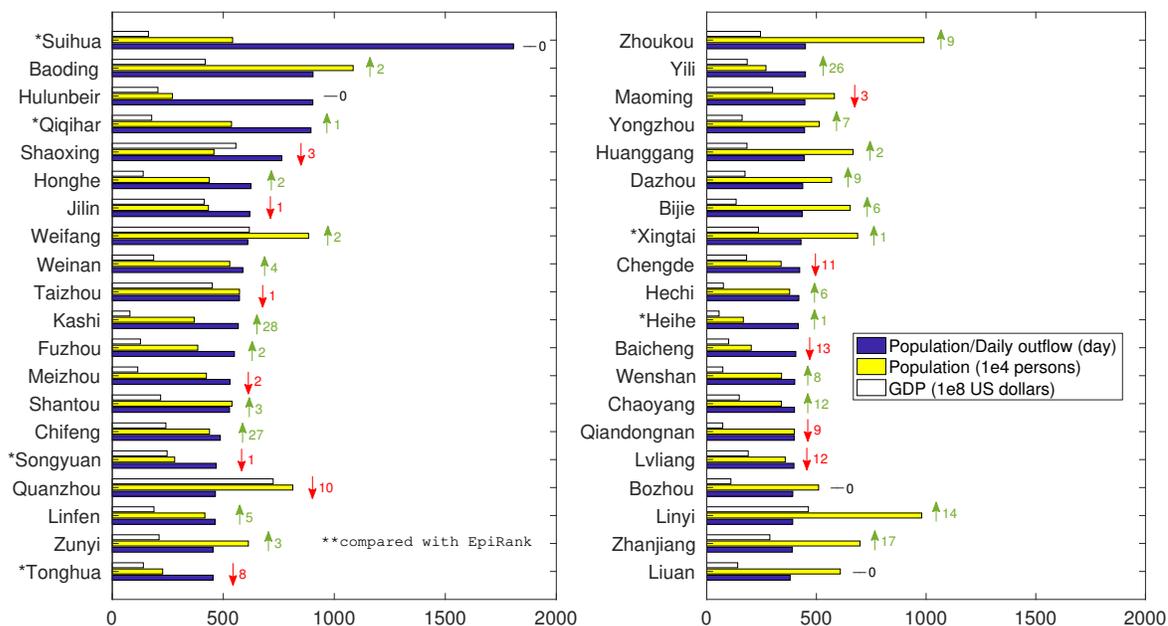}  
\caption{Population/daily outflow as a simplified epidemic hazard indicator (showing top 40 cities). Ranking compared to EpiRank (results in Figure 2); differences marked with green, red or black arrows and numbers. The ratio is effective but less accurate than EpiRank in highlighting the six epicenters (with star marks). Shear population or city GDP does not provide the same indication.} 
\label{Fig3}
\end{figure}

\section*{Discussion}

From the model and results one deduces that, the high epidemic hazard of these small-scale cities computed at low $\alpha$, in which case $h_i$ draws heavily on a city's own infection $U_{ii}$, derives from the combined effect of two factors: a relatively large local population, and a small inter-city transportation flow. Intuitively, a serious epidemic cluster at the regional scale is going to develop, when the region is sufficiently populated, and not much inter-city outflow is dispersing the infection out of the epicenter. This inspires the idea that alternatively, we could compute the population/outflow ratio of each city and use this quantity to indicate cities' epidemic hazard. Results show that (Figure 4) similar to EpiRank, this ratio does serve as a good hazard indicator, under which the six denoted cities are listed with high ranks; furthermore, by contrast, shear population or city GDP, arguably two most considered social-economic indicators of urban regions \citep{Niuet2020}, are not valid to reflect the ground-true ranking. Conceptually, analysis on EpiRank help us pin down these two quantities among various social-economic factors in establishing a promising mathematical explanation of the observed phenomenon. For robust tests, we proportionally increased and decreased values on the flowmaps; results suggest that the effectiveness of this ratio (and certainly the effectiveness of EpiRank) in highlighting the six epicenters is largely invariant to changes in absolute flow strength.

Nevertheless, it is seen that the simple population/outflow ratio, although still effective and easy to compute, is not as accurate and informative as EpiRank. This time the six ground-true cities are overall lower ranked, with only 2 out 6 in top 5$\%$. This is because this ratio only considers a city's own population and transportation condition, whereas EpiRank takes a full account of the regional and then the entire national picture, under the networked dynamics approach. Indeed, it is not empirically inconsistent to argue that the six high-epidemic-risk cities are all located in the north, not only because they themselves have large population and small inter-city outflow, but also because it is exactly that cities in northern China, with which the six cities exchange most outflow population, all tend to have such features and therefore the effect of local clusters is further locked in. The advantage of EpiRank is implied; certainly, the simple ratio is also not able to reveal cities' ability of spreading the disease out, as EpiRank can shed light on with high $\alpha$. 

Although a promising quantitative explanation for the researched phenomenon is developed, it is yet indiscreet to conclude that EpiRank is by any means a sufficiently accurate index of urban epidemic hazards. The current dynamic network model draws little besides the two aspects, urban population and inter-city transportation, and too many real-world factors are left out. Validation of EpiRank results is also difficult to be conducted in a systematic and rigorous way, besides using the six epicenters as the ground truth. Despite a mathematically consistent and empirically effective approach, the proposed simulation framework and the constructed EpiRank index needs further analysis and extensive tests in various settings (e.g., to investigate the situation in the US \citep[e.g.,][]{Changet2021}), before their powers and shortcomings could be substantially uncovered; this study only serves as a first attempt.

\end{document}